# Statistical and mathematical modeling of spatiotemporal dynamics of stem cells

Walter de Back[1,2], Thomas Zerjatke[1], Ingo Roeder[1,3, *]

(1) Institute for Medical Informatics and Biometry, Carl Gustav Carus Faculty of Medicine, TU Dresden
(2) Center for Information Services and High Performance Computing, TU Dresden
(3) Data and Trial Management Unit, National Center for Tumor Diseases (NCT), Partner Site Dresden

## ii. Abstract

Statistical and mathematical modeling are crucial to describe, interpret, compare and predict the behavior of complex biological systems including the organization of hematopoietic stem and progenitor cells in the bone marrow environment. The current prominence of high-resolution and live-cell imaging data provides an unprecedented opportunity to study the spatiotemporal dynamics of these cells within their stem cell niche and learn more about aberrant, but also unperturbed, normal hematopoiesis. However, this requires careful quantitative statistical analysis of the spatial and temporal behavior of cells and the interaction with their microenvironment. Moreover, such quantification is a prerequisite for the construction of hypothesis-driven mathematical models that can provide mechanistic explanations by generating spatiotemporal dynamics that can be directly compared to experimental observations. Here, we provide a brief overview of statistical methods in analyzing spatial distribution of cells, cell motility, cell shapes and cellular genealogies. We also describe cell-based modeling formalisms that allow researchers to simulate emergent behavior in a multicellular system based on a set of hypothesized mechanisms. Together, these methods provide a quantitative workflow for the analytic and synthetic study of the spatiotemporal behavior of hematopoietic stem and progenitor cells.

## 1 Introduction

Despite major advances in the identification of molecular and genetic components and biomarkers of the local bone marrow environments ("niches"), in which hematopoietic stem and progenitor cells (HSPC) reside, much remains to be learned about the spatiotemporal dynamics of normal and aberrant hematopoiesis (1-3). There are many remaining open questions concerning e.g. the localization of HSPC relative to various, potentially different stem cell niches, their chemotactic and migratory behavior within the bone marrow, the heterogeneity of cellular morphologies, the role of niche factors on cell fate decisions, etc.

Modern microscopy provides image data with increasing spatial and temporal resolution, e.g. high resolution imaging of deep tissue of the bone marrow (4), live cell video microscopy of stem cell cultures (5-7), as well as intravital imaging of HSPC within the bone marrow (8-11). However, in order to interpret such rich image data and use it to test scientific hypothesis, e.g. on the spatiotemporal organization of HSPCs, requires statistical and mathematical modeling. On the one hand, data-driven statistical models provide frameworks to describe and quantify experimentally observed aspects of HSPC behavior, to formulate (null-)hypotheses, and to formally compare and potentially distinguish the observed behavior from a formulated hypothesis, such as random behavior. Hypothesis-driven mathematical models, on the other hand, provide mathematical and computational frameworks to test whether a set of assumptions on the cellular behavior and interactions of cells is, in principle, able to generate the observed spatiotemporal regularities.

In this chapter, we provide introductions to common procedures in statistical modeling to quantify (i) spatial distributions, (ii) motility, (iii) cell shape, and (iv) proliferative behavior of cells. In each of these procedures, we specify the types of questions that can be addressed, how to prepare input data, what steps are required to measure statistical properties, how to formulate the null hypothesis, and how to compare observations to this expectation. Moreover, we describe two cell-based modeling frameworks, i.e., the center-based model and the cellular Potts model, in which hypotheses on cellular behavior can be tested computationally. Together, these statistical and mathematical modeling methods provide a full quantitative workflow for the analytic and synthetic study of the spatiotemporal behavior of hematopoietic stem and progenitor cells.

## 2 Methods

### 2.1 Image segmentation and tracking

A prerequisite for statistical and mathematical modeling is the presence of quantitative data. Obtaining these from images is a non-trivial process that typically starts with *image segmentation*, in which an image is partitioned into fore- and background to e.g. separate cells from other structures. If not only static, but also dynamic information is of interest, *cell tracking*, becomes relevant. Here, moving cells are located and linked between image frames and cellular events, such as cell deaths or divisions, can be detected and related to positions of cells in space and time. Both are active fields of research and a wide range of segmentation and tracking methods exists, from manual annotation (12), computer-assisted methods (13-15) to fully automated machine learning algorithms (16,17) . While the presence of segmented images and/or cell tracks is a pre-condition for quantification and statistical modeling, image analysis itself is not in the focus of this publication. Therefore, we refer the reader to available review literature and software packages (see below for a selection of references).

#### 2.1.1 Further reading and software

A brief historical overview of cell segmentation methods, before the advent of machine learning, is given by Meijering (18), while Kan (19) describes some of the opportunities of machine learning in cell image analysis. Caicedo and co-workers (20) present a series of best practices for image acquisition, image processing and subsequent data analysis focusing on high content profiling, while Skylaki et al. (7) provide an overview of the challenges in long-term imaging and quantification.

A number of software packages are available for biological image segmentation, including *Fiji* (21), *Ilastik* (15), *CellProlifer* (22), *CellCognition* (23), and the more recent *fastER* (13). Some of this software also includes features for cell tracking, such as the *TrackMate* plugin (24) included in *Fiji* and the manual and automatic tracking tools in *Ilastik* (15) and *CellCognition* (23). Other software packages are designed specifically for tracking live cells such as *tTt* (12). Wiesmann et al. (25) review a number of additional software tools for quantitative image analysis.

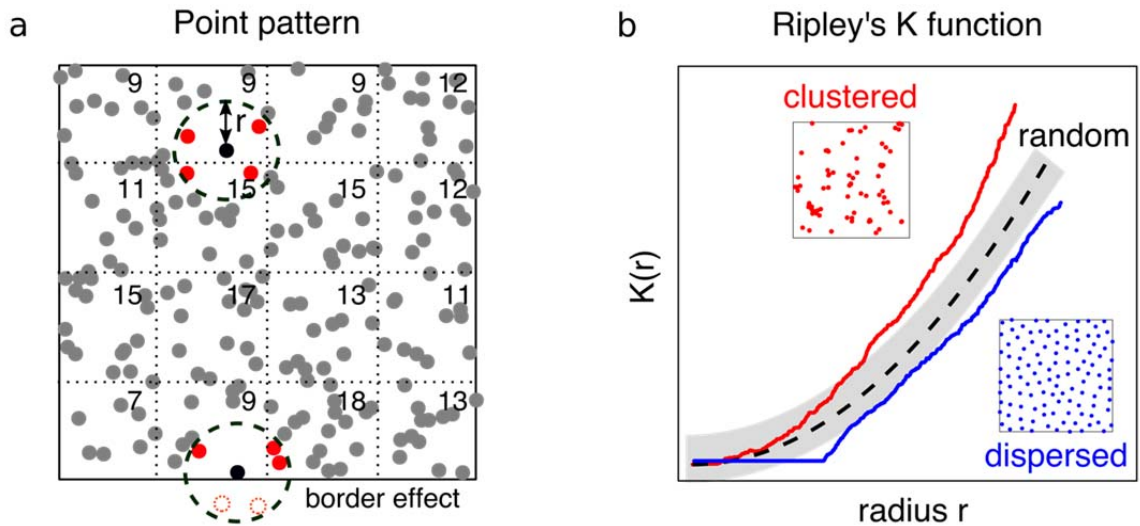

**Figure 1: Spatial distribution:** (a) By representing cellular populations by their centers of mass, their spatial distribution can be studied as a point pattern within an observation area A. To test for homogeneity, the quadrant counting method counts the number of points in sub-regions (dotted lines). To compute the Ripley's K function, one counts for each point, the number of neighboring points within an increasing radius, normalized by the maximum number of pairwise distances, and correcting for border effects where neighboring points cannot be counted. (b) Ripley's K function (aka reduced second moment measure) shows an exponentially increasing function $\boldsymbol{K_{Poisson}(r) = \pi r^2}$ for a 2D completely spatial random (CSR) pattern. Observed K-functions above the CSR expectation indicate clustering, while an observed K-function below the CSR expectation shows that the pattern is regular or dispersed. The grey region indicates the upper and lower global envelopes.

## 2.2 Spatial distribution

Distances between cells are often used as a proxy for cellular interactions. Although we would like to emphasis that local proximity and functional interaction is in general not the same, quantitative information on distances between cells (and potentially other structures) allow to describe patterns of cell motility as well as to identify the presence of (stem cell niche-mediated) local functional or regulatory peculiarities. Therefore, aided by high resolution imaging, many studies focus on measuring distances of stem cells to other cell types and bone marrow structures in order to identify stem cell niches (4,26). Understanding the localization of HSPCs can support answering questions such as: Are hematopoietic stem cells randomly distributed within the bone marrow or does their distribution follow certain structural rules? Are these cells more prominently localized to vascular or endosteum surfaces? Are there signs of attractive or repulsive interactions between different cell types?

Methods from *point pattern analysis*, a statistical framework mostly developed in the context of ecology and geology, can be used to answer these questions. Specifically, this analysis is used to investigate whether a certain distribution of points show signs of regularities. It typically starts by assuming complete spatial randomness (CSR) as a null hypothesis and investigates whether the observed cellular pattern can be described by such a random pattern generating process. CSR suggests that (i) the number of points is proportional to the area of a sub-region (*homogeneity*) and that (ii) the locations of points are independent from each other (*independence*) (see **Note 1**). Based on these assumptions, it is possible to describe the number of points in any given region by a Poisson distribution. To statistically test whether the null-hypothesis of CSR is plausible (in a probabilistic sense), one can use the *quadrant counting* and *Ripley's K function* as described below.

- **Input data:** In point pattern analysis, cells are approximated as point particles. These are given by the 2D or 3D coordinates of the center of mass of the segmented cell or, alternatively, by manually tracking. The cell density is described by the *intensity* of the point pattern and can be estimated by $\lambda = \frac{n(x)}{|W|}$ where $\lambda$ is the intensity, assuming the point process is homogeneous, $x$ is the set of point pattern of size $n(x)$ and $|W|$ is the size of the observation window.
- **Quantify and testing homogeneity**: To test for homogeneity of the point pattern, one can check whether the regions of equal area contain roughly equal numbers of points (see Figure 1a), i.e. applying *quadrant counting*. To test the null-hypothesis of CSR, a typical $\chi^2$(chi-squared) statistic can be used:

$$\chi^2 = \sum_j \frac{(observed - expected)^2}{expected} = \sum_j \frac{\left(n_j - \frac{n}{m}\right)^2}{\frac{n}{m}} \tag{1}$$

where $n_j$ is the number of points observed in subregion $j$, $n$ is the total number of points in the observation window and $m$ is the number of (equally sized) subregions. Under the CSR null hypothesis, the test statistic is approximately $\chi^2$-distributed with $m - 1$ degrees of freedom.

- **Quantify spatial correlation:** Independence of the localization of points can be violated either by (i) *clustering* of points, i.e. points are closer together than expected, e.g. due to some attractive interaction or by (ii) *dispersion*, i.e. points are farther apart than expected, e.g. due to a repelling activity. A commonly used method to analyze spatial correlation is the *reduced second moment measure*, better known as *Ripley's K-function* (27). Ripley's K-function (to be estimated by the empirical K-function given by equation (2)) counts the average number of neighboring points within a certain radius (see Fehler! Verweisquelle konnte nicht gefunden werden.a), normalized to the possible number of pairwise distances $n(n - 1)$ within area $A$

$$\widehat{K}(r) = \frac{A}{n(n-1)} \sum_{i=1}^{n} \sum_{j=1, j \neq i}^{n} k_{ij}(r) e_{ij}(r) \tag{2}$$

where $A$ is the observation area, $n$ is the observed number of points and $k_{ij}$ indicates whether the pairwise distance $d_{ij}$ is smaller than $r$:

$$k_{ij}(r) = \begin{cases} 1 & if\ d_{ij} \leq r, \\ 0 & otherwise \end{cases} \tag{3}$$

The edge correction factor $e_{ij}(r)$ compensates for the fact that we cannot count points that lie within the radius $r$ but outside of the observation window (see Fehler! Verweisquelle konnte nicht gefunden werden.a). Therefore, without correction, the number of points is underestimated for larger radii. Several correction methods exists, ranging from simple border methods that suffice for large data set to more complex methods such as isotropic or translation correction that are advised for smaller data sets (28).

- **Comparison to CSR:** From the homogeneity and independence properties of CSR it follows that the point pattern can be described as the realization of a *Poisson process*. This can be used to calculate the expected number of points lying within a distance $r$ of a typical random point. In 2D, under the CSR assumption, the theoretical K-function is $K_{Poisson}(r) = \pi r^2$ and in 3D it is given by $K_{Poisson}(r) = \frac{4}{3}\pi r^3$. The empirical and theoretical K-functions are typically compared graphically (see Figure 1b). An empirical K-function that is above the theoretical CSR expectation, i.e. having more than expected neighboring points, indicates a clustered pattern. Conversely, a line below the theoretical expectation, i.e. having less than expected neighbors at a certain radius, is a sign of regularity or dispersion (see **Notes 2 and 3**).
- **Significance testing of CSR**: To test whether the empirical (i.e. observed) K-function deviates statistically significant from the expected K-function (under CSR), one can perform a *Monte Carlo test* where the K-function under the CSR null-hypothesis is repeatedly simulated. For this, the maximum and minimum deviation of the simulated and the theoretical K-function values are used as a global envelop (see Figure 1b and **Note 4**). The observed clustering or dispersion is considered significantly different from CSR when the observed K-function crosses out of the global envelope. The significance level associated with this *Monte Carlo* test is determined by the number of simulations used to calculate the global envelope, i.e. $\alpha = 1/m + 1$ where $m$ is the number of simulations. I.e., for $m = 19$ simulations a test at the typical significance level of $\alpha = 0.05$ would be obtained (see section 10.6 in (29)).

### 2.2.1 Further reading and software

In case the above analysis shows signs of inhomogeneous point pattern, more analysis is required, e.g. to reveal dependence of the intensity of the point pattern to other factors. *Spatial covariate analysis* can be used to analyze whether cell density is correlated with covariates such as the concentration of a chemokine or the distance to the vascular surface (chapter 6.6 in (29)). When localization data on multiple cell types in the same culture or tissue are available, one might ask whether there is an interaction between these cell types. This requires a more elaborate *multitype point pattern* analysis, where one tests the null hypothesis of *complete spatial randomness and independence* (CSRI) e.g. by computing empirical K-functions in a pairwise fashion to reveal attraction or repulsion of cell types around or away from each other (chapter 14 in (29)). Suggested textbooks on statistics for spatial data are (29-31). The latter accompanies the key software package for point pattern analysis `spatstat` (http://spatstat.org) for the statistical software package *R* which includes functions for the methods mentioned above, among many others. For *python*, some functions for e.g. computing the (edge corrected) K-function are available in `astropy` (http://astropy.org).

## 2.3 Cell motility

Cell migration is a fundamental property of cells that changes in response to chemical, mechanical and genetic perturbations. Disruption of normal cell motility is potentially associated with disease, e.g. due to immunological response or malignant processes. Unravelling such anomalies requires a rigorous analysis of cell migration patterns.

Statistical analysis of cell motility allows answering questions such as: How fast do cells migrate? Do cells migrate persistently or are the cells constrained, e.g. by a crowded environment? Is there evidence of directionality in the migration patterns?

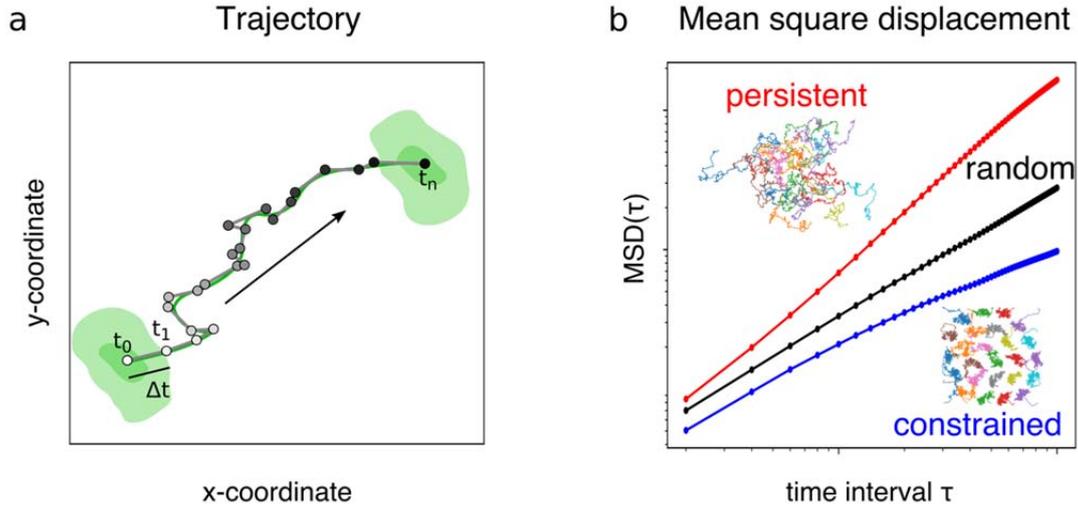

**Figure 2: Cell motility**. (a) The trajectories of cells measured over regular time intervals can be studied by investigating how the displacement of the cell relates to the time interval, assuming the process is time-invariant. This can be quantified by the mean square displacement, calculated for each individual cell or averaged over the population. (b) The mean square displacement is shown on a log-log plot and can be described by the anomalous diffusion model $\boldsymbol{MSD = D\tau^\alpha}$. For a normal diffusion (random) process, the displacement is expected to be proportional to the time interval, and appears as a straight line and have anomality parameter $\boldsymbol{\alpha = 1}$. Cell trajectories above this expectation with $\boldsymbol{\alpha > 1}$ move faster-than-diffusion (superdiffusive), indicating persistent movement. Cell below the expectation with $\boldsymbol{\alpha < 1}$ move closer-than-diffusion (subdiffusive) which indicates constrained movement, often encountered in crowded environments.

The most basic statistics about a cell's trajectory are its length and its curviness or tortuosity. However, to characterize the type of motility a cell exhibits, more elaborate analysis is required. Analogously to the CSR in analyzing spatial distributions of cells, here, we assume randomness as a null hypothesis. Under this assumption, the amount of space that a particle "explores" is proportional to the time interval, which can be measured by calculating the mean squared displacement (MSD). The MSD relates the mean displacement of a particle to different time intervals.

The most common model to describe deviations from a complete-random cell motility is the *persistent random walk* (PRW) model, which account for a degree of persistence of motion. While this model has been shown to accurately describe cell migration on 2D surfaces (32), recent studies have found that it fails to describe cell motility in more biologically relevant 3D environments (33). Although the PRW model can account for faster-than-diffusive (super-diffusive) persistent motion, it cannot describe slower-than-diffusive (sub-diffusive) motility, which is commonly found in cells in confined environments including porous media such as trabecular bone (34). In this case, the general *anomalous diffusion* (AD) model, which relates the cell displacement to time interval with a simple power law, $MSD_{AD}(\tau) = D\tau^\alpha$, is more appropriate to describe both population-average and individual cell migration paths.

- **Input data:** Cell trajectories should be recorded in the form of a table with columns: cell ID, time, and x, y and z coordinates of the cell centroid in consistent units (typically minutes and microns) and equally spaced time intervals (see **Notes 5 and 6**).

- **Quantify length and tortuosity of trajectory**: The most basic statistic about the trajectory is its length. The total length of a trajectory is simply the sum of the trajectory segments $L = \sum_{t=0}^{t_{max}} \sqrt{(x_t - x_{t+1})^2 + (y_t - y_{t+1})^2}$. The curviness or tortuosity of the trajectory can also be calculated in a straightforward fashion as the *arc over chord ratio*: the ratio between total length of the trajectory $L$ and the length of a straight line between start and end points: $T_{AOC} = L/L_X$ where $L_X = \sqrt{(x_{t_{max}} - x_0)^2 + (y_{t_{max}} - y_0)^2}$. However, this measure fails to capture differences between long curved trajectories and short twisted ones because it quantifies the global deviation of the trajectory from a straight line and does not account for local measures such as curvature. A more robust measure of tortuosity calculates the *mean of the local curvatures* along to the trajectory:

$$T_C = \frac{1}{N} \sum_{t=0}^{t_{max}} \frac{x'(t)y''(t) - y'(t)x''(t)}{[x'(t)^2 + y'(t)^2]^{3/2}} \quad (4)$$

where $x'(t)$ and $x''(t)$ denote the first and second derivative of the spline approximating the trajectory (see Figure 2a). Alternatively, if interested mainly in the cell's persistence, one can compute the *direction autocorrelation* (see section 2.3.1).

- **Test for time invariance**: Quantities such as the MSD assume time-invariance of the trajectories, i.e. the velocities should be approximately constant over the course of the experiments. To test this, one first calculates the displacements for each cell and for every time point $dx = x(t + \tau) - x(t)$ and $dy = y(t + \tau) - y(t)$, where $\tau$ is the time interval (typically simply 1). Then take the mean and standard deviation of the velocity over all cells $v_t = \langle \sqrt{(dx^2 - dy^2)}/\tau \rangle$. As a graphical test, one can plot the mean velocities and their standard deviations over time (see **Note 7**). If both remain approximately constant over time, they can be considered as being time-invariant (see **Note 8**). As a formal test, one can perform a linear regression on the mean velocity (see **Note 9**). A time-dependency is indicated by a slope deviating from zero, which can be statistically confirmed by testing the corresponding null-hypothesis (i.e., slope = 0) and observing a p-value smaller than a predetermined significance level (e.g. $\alpha = 0.05$).

- **Compute mean square displacement**: The mean square displacement (MSD) of individual trajectories is computed by calculating the mean displacement for different time intervals (here for 2D):

$$MSD_{cell}(\tau) = \frac{\Delta t}{t_{max} - \tau} \sum_{t=0}^{t_{max}-\tau} \left( \left(x(t+\tau) - x(t)\right)^2 + \left(y(t+\tau) - y(t)\right)^2 \right) \quad (5)$$

where with $\tau = n\Delta t$ with $n = 1, 2, \dots, n_{max}$ and $\Delta t$ is the time between individual frames (see **Note 10** and Figure 2a). The MSD of the whole population of cell trajectories, called aggregate or ensemble MSD, is calculated by averaging over the MSD for all cells at each particular $\tau$:

$$MSD_{ensemble}(\tau) = \frac{1}{n_{max}} \sum_{n=1}^{n_{max}} MSD_{cell\ n}(\tau) \qquad (6)$$

- **Fit anomalous diffusion model**: To estimate the parameters of the anomalous diffusion (AD) model, i.e. to estimate the diffusion coefficient $D$ and the anomaly parameter $\alpha$, we fit the MSD of a cell or ensemble using a nonlinear regression (see **Note 11**). Importantly, only the first 20-30% part of the data (containing the smallest time intervals) should be used in the fitting procedure (see **Note 12**). For a normal diffusion (random) process, the displacement is expected to be proportional to the time interval, and appears as a straight line and has anomaly parameter $\alpha = 1$. An MSD with $\alpha > 1$ exhibits faster-than-diffusion motion (superdiffusion) which indicates persistence. In contrast, an MSD with an estimated $\alpha < 1$ move slower-than-diffusion (subdiffusion) which indicates constrained movement that is often associated with crowded environments.
- **Significance testing**: To test whether the fitted AD model is significantly different from the random expectation, one can apply *Monte Carlo testing* as described in section 2.2 by fitting the AD model to the MSD of a number of simulated random walk trajectories.

### 2.3.1 Further reading and software

Further analysis can consist of spatial autocorrelation metrics, such as the velocity and direction autocorrelation, which measure how a quantity correlates with itself over different time scales. For instance, to compare the persistence of cells between conditions, the *direction autocorrelation* can be calculated, e.g. with the Excel-based *DiPer* software tool (35). With such analysis, one is able to better distinguish between the common models for cell migration, the PRW, anisotropic PRW (36), anomalous diffusion (37), and fractional diffusion models (38). An excellent review on computational methods for measuring cell migration is given in (39).

Several software packages offer tools to analysis cell trajectories, including *CellProlifer Tracer* (http://cellprofiler.org/tracer), *MotilityLab*, a website as well as an *R* package (http://www.motilitylab.net), the *python* package *TrackPy* (https://soft-matter.github.io/trackpy) and *TrackMate*, an *ImageJ* plugin (https://imagej.net/TrackMate).

## 2.4 Cell shape analysis

Cell shape is one of the most common properties used to characterize cellular phenotypes, in particular in high-content imaging and drug screening. This is due to the fact that cell morphology can be used as a proxy for a range of cellular processes including cell death, division, polarity and motility. Moreover, the shape of a cell is a relatively easy accessible property given appropriate cytoplasmic staining. Statistical analysis of cell shape aims to answer questions such as: What is the source of heterogeneity in cell shape within a population? Can one identify subpopulations with similar cell shapes? Are cell shapes correlated with cell lineages?

Here, we present a procedure to explore the heterogeneity of cell shapes in which shapes, represented by regions in binary masks, are quantified using a number of 2D shape descriptors. Subsequently, differences within the population and the potential presence of subpopulations are analyzed and visualized by creating a "shape space" using principal component analysis.

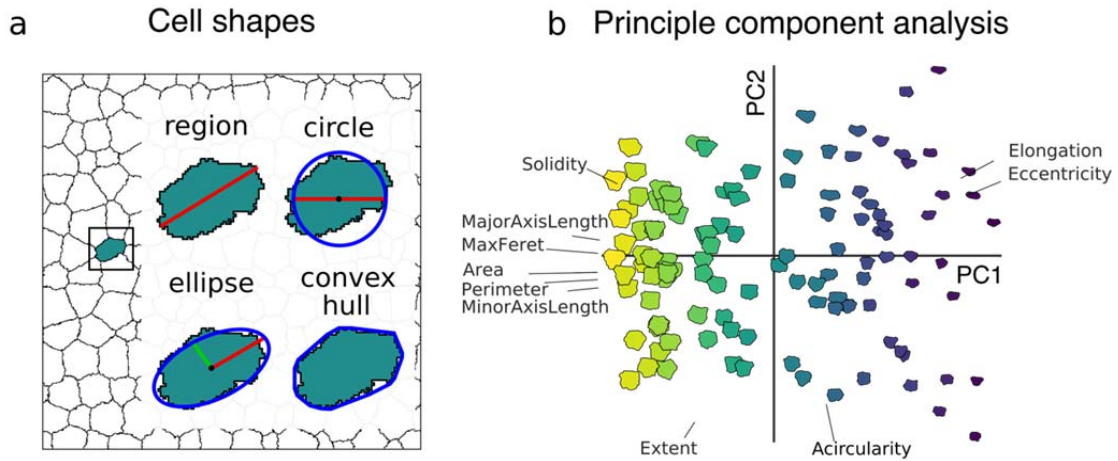

**Figure 3: Cell shape analysis**. (a) After segmentation, cell shapes can be quantified according to a large number of shape descriptors called region properties, ranging from simple quantities such as area and perimeter to more complex ones based on various geometric models such as the circular, elliptic approximations as well as the convex hull. (b) To explore the variance in a population of cell shapes described by shape descriptor, one can apply a dimensionality reduction method such as principal component analysis (PCA) to explore the distribution of cell shape and investigate the most distinctive descriptors (feature selection).

- **Input data**: The input for cell shape analysis consists of a set of binary images in which cells are separated and segmented from the background (see section 2.1). If multiple cells are present in an image, each cell must be uniquely identified using *connected-component labelling*, which detects and uniquely labels connected regions in an image, after which a separate binary image can be generated for each cell.
- **Region-based statistics**: Each 2D labelled region can be described using a wide range of statistics. These statistics are either be based on the region itself, or on one of several geometrical approximations of the region. Some of the most commonly used shape statistics are listed below:
    - The region itself can be quantified in terms of the *area A*, i.e. the number of pixels that form the area A and the *perimeter P*, i.e. the length of a line through the centers of the border pixels, or the *maximum Feret* (a.k.a. *caliper*) diameter as the longest distance between two points in the region.
    - Based on a **circular approximation**, one can quantify the *equivalent diameter* or a circle with the same area $= \sqrt{4A/\pi}$. The *acircularity* is defined as the ratio between the observed perimeter and the expected perimeter of a circle, $a = P^2/4A\pi$.
    - By calculating the **elliptic approximation** (using the moments of inertia), one can measure the length of the major (long) and minor (short) axes and define the *elongation* as the ratio between the two, quantify the shape's *orientation* as the angle between the x-axis and the major axis and measure the *eccentricity*, defined as the ratio between the two focal points over the major axis length.

- o   Properties related to the **bounding box**, the minimal rectangle containing all the points of the region, are the *extent* (a.k.a. *rectangularity*), the ratio between the area of the region and its bounding box and the *aspect ratio*, the ratio between the width and height of the bounding box, but these should be handled with care (see **Note 14**).
- o   The **convex hull** of a region is the smallest convex polygon (non-intersecting polygon with all interior angles less than 180°) that contains all points of the region and can be used to calculate the area $A_{hull}$ and perimeter $P_{hull}$ of the convex hull. Based on those, one can calculate the *convexity* as the ratio between the perimeters of the convex hull and the original region $c = P_{hull}/P$, and the *solidity*, the ratio between the areas $s = A/A_{hull}$, where the area of the convex hull is equal or larger than the regions area.

- **Feature selection and dimensionality reduction**: After quantifying each cell shape using the descriptors above, a matrix is obtained with rows containing samples (shapes) and columns corresponding to a specific shape descriptor, often containing more than 20 features per cell shape. However, many of these shape descriptors quantify similar aspects of the shape, i.e. there may be strong correlations between descriptors. Moreover, depending on the most relevant shape properties in a particular study, some descriptors may be redundant. To reduce the dimensionality of the data and find the most relevant features, one can apply e.g. a *principal component analysis* (PCA). PCA allows to represent the high-dimensional data in a lower-dimensional (typically 2D or 3D) "shape space" that still captures most of the variance in the data (see Figure 3b and **Note 15**). To interpret the principal components in terms of the descriptors, one can look at the *component loadings*, which describe the correlation between descriptors and the principal components.

### 2.4.1   Further reading and software

Apart from binary masks, cell shapes can also be represented by distance maps or polygonal outlines and subjected to different encodings such as Fourier, elliptic Fourier (40) and Zernike decompositions. Pincus and Theriot (41) present an extensive review and quantitative comparison of these methods. A number of studies focus on the dynamics of cell shape rather than static shapes, e.g. by using time series of shape descriptors (42) or modeling trajectories in shape space (43).

A number of software packages exist to extract region-based statistics from binary image masks, including `regionprops` in the image processing toolbox in *Matlab*, and a function by the same name in the `skimage.measure` *python* package. Implementation for PCA are offered as *pca* in the statistics and machine learning toolbox in *Matlab* and as in the `sklearn.decomposition.PCA` package in *python*.

## 2.5   Cell lineage tree analysis

Cellular genealogies, also denoted as cell lineage trees, are pedigree-like structures that represent the complete divisional history of a cell (Figure 4a). Analyzing these structures can help answering questions like: Are certain cellular characteristics inherited upon division and if so, how far-reaching are these correlations? Do certain events, e.g. cell death, occur preferentially in the genealogical vicinity of other such events? When does the decision to differentiate take place? Is a differentiation process regulated in an instructive (i.e. by differential differentiation propensities) or selective manner (i.e. by differential death rates)?

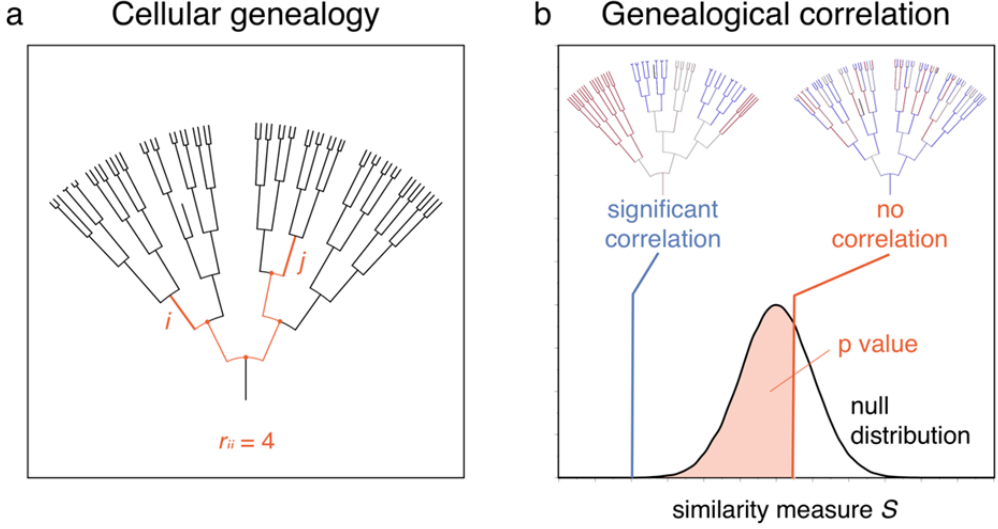

**Figure 4: Lineage tree analysis.** (a) Cellular genealogies (also denoted as cell lineage trees) represent the complete divisional history of a founder cell (at the bottom). Cells are depicted as straight lines with the length corresponding to the cell's lifetime. Upon division, a mother cell is connected to its two daughter cells. The topological distance $r_{ij}$ between two cells $i$ and $j$ is defined as the number of divisions that separate the cells within the tree. (b) In order to detect genealogical correlations of a feature of interest, e.g. the length of cell cycle (as color-coded here within the genealogies), a similarity measure is calculated and compared to a null distribution of randomly expected values. A small p value (defined as the fraction of values generated under the null hypothesis that are less or equal to the actually observed value) indicates a clustering of the feature of interest (left genealogy), while a large p value indicates a random distribution of the feature of interest (right genealogy.)

Genealogical information can be used to detect correlations of discrete events, e.g. cell death or onset of differentiation marker, or continuous (i.e. metric) features, e.g. cell cycle length or cell motility (44,45) (see **Note 16**). Four steps are necessary to detect potential correlation structures: (i) represent the genealogical information in an efficient way, (ii) define a similarity measure that reflects genealogical correlations, (iii) find an appropriate null model that reflects the correlation measure's expected distribution without any correlation, and finally (iv) compare the actual value with this null distribution.

- **Input data**: An efficient representation of the genealogical structure is a cell numbering according to the following rule: For a mother cell $i$, label its two daughter cells as *2i* and *2i+1*. Using this numbering, it is easy to calculate a cell's generation as well as the topological distance $r_{ij}$ between a pair of cells $i$ and $j$, i.e. the number of divisions that separate those events within the genealogy (Figure 4a).

$$generation(i) = \lfloor log_2 i \rfloor$$
$$r_{ij} = \lfloor log_2 i \rfloor + \lfloor log_2 j \rfloor - 2\lfloor log_2 c(i,j) \rfloor + \begin{cases} 1 & if\ c(i,j) = i\ or\ j \\ 0 & otherwise \end{cases}, \quad (7)$$

with *c(i,j)* being the last common ancestor of cells *i* and *j*, which is calculated by iteratively determining the ancestors of both cells. $\lfloor \cdot \rfloor$ here denotes rounding to the nearest lower integer.

- **Quantify genealogical correlations**: Depending on the outcome of interest, i.e. a discrete event or a continuous feature, two different similarity measures can be defined. Both are based on the topological distance $r_{ij}$ between a pair of cells $i$ and $j$.

- **Discrete events**, e.g. cell death or onset of markers: For every event *i* in the genealogy, the minimal distance to the closest other event is determined. This is averaged over the set of all events *E*:

  - $S = \frac{1}{|E|} \sum_{i \in E} \min r_{ij}$ (8)

- **Continuous quantities**, e.g. motility or cell cycle length: For every pair of cells the absolute difference of the feature of interest *m* is determined and weighted according to their topological distance. This is averaged over the number of pairs:

$$S = \frac{1}{n^2} \sum_{i,j} \frac{|m_i - m_j|}{2^{r_{ij}}} \quad (9)$$

To especially estimate the range of correlation structures, the measure can be restricted to pairs of cells that have the same topological distance *k*:

$$S^{(k)} = \frac{1}{n^2} \sum_{r_{ij}=k} |m_i - m_j| \quad (10)$$

- **Null model for randomly expected distribution**: The defined similarity measures are only meaningful relative to a "neutral" control, i.e. considering what would be expected without any correlation. To determine this so-called null distribution, a permutation procedure is suggested that randomizes the correlation structure. There are two general ways of randomization: either (a) by reassigning the attribution of mother and daughter cells or (b) by reassigning the feature of interest. The randomization procedure is repeated a sufficiently large number of times (e.g. 100,000 times), each time calculating the chosen correlation measure. This permutation procedure then gives the (empirical) null distribution (Figure 4b).
  - **Randomize mother-daughter attribution**: For each generation, the assignment of cells to mother cells of the previous generation is randomized. This randomization changes the genealogical topology and thus makes it necessary to recalculate topological distances of cell pairs for each permutation step (see **Note 17**).
  - **Randomize feature of interest**: Using this randomization procedure, the topology of the genealogy is not changed, but the feature of interest, e.g. cell motility, is reassigned to the cells. This reassignment should be restricted to cell of the same generation to avoid intermingling with general temporal effects, i.e. changes of the feature over time that are independent of genealogical correlations.
- **Compare similarity measure to null distribution**: The actual value of the similarity measure *S* is now compared to this null distribution. An empirical p-value can be computed to decide whether there is a genealogical correlation of the feature of interest. This p-value is defined as the fraction of values of the similarity measure generated under the null hypothesis that are less or equal to the actual value (Figure 4b).

A large p-value would indicate that the actual value is within the range of values that are expected (i.e. rather likely) without any correlation Figure 4b, right genealogy), while a small p-value indicates a low probability to observe such a genealogy without any correlation, i.e. suggesting a clustering of the feature of interest that is beyond the randomly expected (Figure 4b, left genealogy). Formal testing can be done by comparing the empirical p-value with a predefined significance level.

Restricting the measure to cell pairs that have the same topological distance $k$ (see eq. (9)) and repeating the analysis for a range of values of $k$ will provide information about how far-reaching correlations are.

#### 2.5.1 Further reading and software

The statistical approach described above had already been used to disentangle differentiation process in a hematopoietic cell line (45). An approach to cluster cell lineage trees according to common patterns and defining centroid trees that represent the characteristic patterns can be found at (46). Another statistical method to discriminate distinct groups of genealogies is described in (47). Branching process models can be used to estimate differentiation rates from genealogical data (48,49). A method to infer and discriminate different spatiotemporal effects on cell state transitions is described in (50). In (51) the authors provide a way to fit stochastic models to lineage trees. To our knowledge, there is no publicly available software implementing methods to analyze cellular genealogies, unfortunately.

## 2.6 Cell-based modeling

The above (data-driven) statistical modeling provides a framework to analyze (image-based) experimental data in order to reveal patterns, regularities and interactions in distribution, motility, shapes and/or genealogies that indicate the presence of regulatory mechanisms. However, statistical modeling does not necessarily provide enough information to pinpoint what mechanism may be responsible for the observed regularities. At this point, hypothesis-driven mathematical modeling can be very useful. Such approaches provide formal mathematical or computational frameworks in which one can formulate hypotheses on the suspected responsible mechanisms in mathematical expressions that can be simulated in a computer. This process forces one (i) to make all assumptions explicit, (ii) to formulate the proposed mechanisms in an unambiguous fashion, and (iii) to provide a complete framework including e.g. the presence of unknown factors (as free parameters) and stochastic processes (as random variables). Such models are *generative* in the sense that they generate artificial data, based on a set of hypotheses and corresponding assumptions. A consistent model will include parameters that are biologically meaningful, interpretable, and, under appropriate parameter choices, be able to generate the same patterns or regularities as observed in the experimental system, under normal conditions as well as perturbations. However, even in such an ideal case, i.e. having identified a model that is consistent with observed data, one must interpret results with care. In principle, from such a consistency one can only conclude that the model's assumptions on regulatory mechanisms are *sufficient* to generate the observed pattern, but one cannot necessarily exclude other mechanisms of regulation. In contrast, if a model description does not allow to consistently describe / explain the data under consideration, one can exclude this set of hypothesis / assumptions as potential explanations (i.e. *falsification* strategy).

One can distinguish two general approaches to mathematical and computational modeling of biological populations: cellular populations are either described as continuous or as discrete quantities. *Continuous models*, implemented e.g. by ordinary or partial differential equations, have the advantage that they, when formulated in simple terms, allow rigorous mathematical analysis that can reveal general relationships between processes. However, continuous models are generally not well-suited to capture heterogeneity in populations or to predict the behavior of small populations where stochastic effects might play a dominant role. Therefore, many modeling studies, in particular those concerning stem cell behavior (e.g. in the intestinal crypt or in the bone marrow niches), have used *discrete models* instead. In a discrete approach, cells are modelled as individual objects with certain properties and a set of (local) interaction rules, using so-called *agent-based* or, more specifically *cell-based models* (52). Since these models are often too complex to allow rigorous mathematical analysis, they are typically simulated by a computer and are hence often denoted as *computational models.* A number of such models have been proposed in the area of hematopoietic stem cell organization (53,54), but these did not explicitly account for spatial aspects including spatial distribution or cell migration.

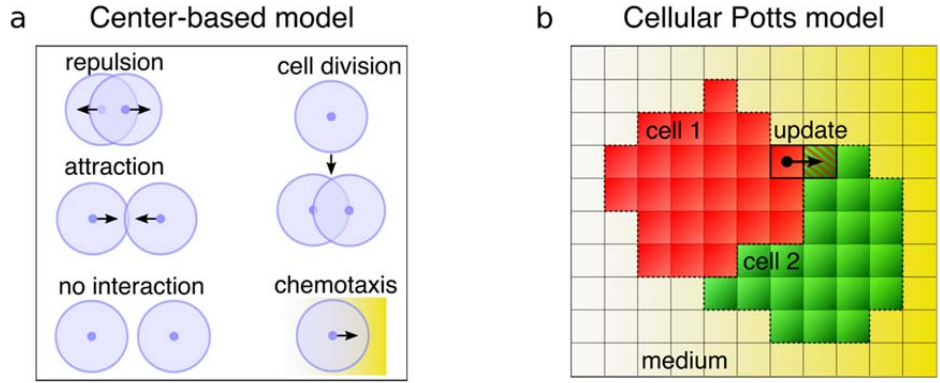

**Figure 5: Cell-based modeling.** (a) In the center-based model, cells are modeled by their centers in continuous space and forces are applied according to the distances from their neighbors: a repelling force when cells are too close (volume exclusion) and an attractive force when cells are in intermediate distance (cell-cell adhesion). Cells can also divide by placing daughter cells according to a random unit vector and can respond to external fields of chemokine gradients. (b) In the cellular Potts model, cells are modeled on as domains on a lattice and are governed by an energy function with terms to constraint volume and perimeter as well as cell-cell surface lengths. By randomly attempting to copy states between neighboring lattice sites, the cells change shape and move in order to minimize the energy.

For spatial cell-based modeling that allows these aspects to be explicitly represented and studied, several well-established theoretical frameworks exist. The most common methods include *cellular automata* (CA) models (55), *cellular Potts models* (CPM) (56), *center-based models* (CBM) (57), *vertex models* (VM) (58,59) and *subcellular element models* (SEM) (60), ordered by increasing complexity. Within each of these frameworks, it is possible to simulate self-organization in tissues and heterogeneous populations with cellular processes such as motility, interactions, division and death (61). However, there are also notable differences. One key difference is in how cells are represented spatially. Whereas some (*on-lattice*) methods use a discretized space (CA, CPM), others (CBM, VM, SEM) use continuous space (*off-lattice*). Some methods treat cells as point particles without explicit volume (CA, CBM), while others model cells with explicit shapes (CPM, VM, SEM). Another key difference is in how *model dynamics* are generated. While CA models are rule-based, CPM and some VM models formulate an energy function that is minimized, and in other methods, dynamics are generated by explicit calculation of forces. Here, we describe two of these methods that represent a cross-section of these difference:

the center-based model (off-lattice, particle-like, force-based dynamics; Figure 5a) and the cellular Potts model (on-lattice, explicit cell shape, energy-based dynamics; Figure 5b).

- **Center-based model (CBM)**: In center-based models (following the notation in (61)), cells are represented as particle-like objects defined by their centers (see Figure 5a), modelled as a set of points $\{\mathbf{x_1}, ..., \mathbf{x_N}\}$, that are free to move in space. Each cell has a radius $R$ and two cells are assumed to be neighbors if their centers are close $\|\mathbf{x}_i - \mathbf{x}_j\| = \|\mathbf{r}_{ij}(t)\| < r_{max}$ where $r_{max}$ is the interaction radius. We can then specify a *repelling force* between cells if they are closer than a natural separation distance $\|\mathbf{r}_{ij}(t)\| < s_{ij}$ to model volume exclusion and an *attractive force* if the distance between cells is between the natural separation and interaction radius $s_{ij} \leq \|\mathbf{r}_{ij}(t)\| \leq r_{max}$ to model cell-cell adhesion and *no interaction force* in case the distance between cells is larger than the interaction radius $\|\mathbf{r}_{ij}(t)\| > r_{max}$:

$$\mathbf{F}_{ij}(t) = \begin{cases} \mu_{ij} s_{ij} \hat{\mathbf{r}}_{ij}(t) \log\left(1 + \frac{\|\mathbf{r}_{ij}(t)\| - s_{ij}}{s_{ij}}\right), & \text{for } \|\mathbf{r}_{ij}(t)\| < s_{ij} \\ \mu_{ij}(\|\mathbf{r}_{ij}(t)\| - s_{ij}) \hat{\mathbf{r}}_{ij}(t) e^{-k_c \frac{\|\mathbf{r}_{ij}(t)\| - s_{ij}}{s_{ij}}}, & \text{for } s_{ij} \leq \|\mathbf{r}_{ij}(t)\| \leq r_{max} \\ 0, & \text{for } \|\mathbf{r}_{ij}(t)\| > r_{max} \end{cases} \quad (11)$$

where $\mu_{ij}$ is a spring constant controlling the size of the force, depending on the interacting cell types, $\hat{\mathbf{r}}_{ij}$ is the unit vector giving the force direction and $k_c$ is a parameter controlling how the attractive force decays with distance between centers.

To compute the dynamics for this model, for simplicity, we assume all cells have identical mechanical properties and use force balance to calculate the model equation. The new position of each cell can then be calculated by the sum of the forces acting on the cell:

$$\mathbf{x}_i(t + \Delta t) = \mathbf{x}_i + \frac{\Delta t}{\eta} \sum_{j \in \mathcal{N}_i(t)} \mathbf{F}_{ij}(t) \quad (12)$$

where $\mathbf{F}_{ij}(t)$ is the current force between cell $i$ and $j$, $\mathcal{N}_i(t)$ is the set of neighboring cells of cell $i$, $\eta$ is the damping constant and $\Delta t$ is the time step which must be selected appropriately small to guarantee numerical stability.

- **Cellular Potts model (CPM)**: In the cellular Potts model, each cell is spatially represented by a domain on a lattice (Figure 5b). Instead of calculating forces directly, forces are encoded implicitly by an energy function that provides constraints for cell shape and motility. In particular, the energy function consists of multiple terms that represent the interaction energy, accounting for adhesive interactions between cell types $\tau_i$ and $\tau_j$, and the area and perimeter constraints, which penalize deviations of the actual cell area $a$ and perimeter $p$ from a target area $A_T$ and target perimeter constraint $P_T$ respectively:

$$E = \sum_{interfaces\ i,j} J_{\tau(\sigma_i),\tau(\sigma_j)} + \sum_{cells,\sigma>0} \lambda_A (a - A_T)^2 + \sum_{cells,\sigma>0} \lambda_P (p - P_T)^2 \qquad (13)$$

where the $\lambda$ terms act as weights on the different terms.

Motility is modeled by random sampling of the lattice (modified Metropolis algorithm) whereby, for every sampled lattice point $x_S$, we evaluate the change in energy $\Delta E$ if we would copy the state of the lattice to a randomly sampled neighboring site $x_T$. The probability of actually performing the update depends on the change of energy:

$$P(\Delta E_{x_S \to x_T}) = \begin{cases} 1 & for\ \Delta E \leq 0 \\ e^{-\frac{\Delta E}{T}} & otherwise \end{cases} \qquad (14)$$

where proposed updates that decrease energy $E$ are always accepted, while updates that increase energy are only accepted with a (Boltzmann) probability that decreases with the energy difference. The parameter $T$ can be used to modulate this probability decrease and models the amount of allowed membrane fluctuations. Effects of external signals such as chemoattractant can be incorporated into the CPM by adding a chemotaxis energy:

$$\Delta E_{x_S \to x_T} = \lambda_C (c_{x_T} - c_{x_S}) \qquad (15)$$

Both cell-based models described above are generic frameworks that can be used to simulate the dynamics of multicellular systems such as the behavior of hematopoietic stem cells *in vitro* or *in vivo*. Basic cellular behavioral mechanisms such as cell motility, chemotaxis and cell division can be simulated in isolation or in combination to investigate the type and variety of emergent patterns these give rise to. In particular, the results of the statistical modeling described above, formulated as a set of assumptions and parameters on cellular behavior and interactions, can be formalized and simulated under different sets of conditions, *ceteris paribus*. Moreover, by recording the spatial locations, cell trajectories, cell shape parameters and divisional history of the simulated cells, one can perform the same statistical analysis on the simulated data and on the experimental data, thus allowing for a direct quantitative comparison and, therefore, check for consistency.

There are several subtle but key differences in the manner in which basic cellular mechanisms such as motility, adhesion, chemotaxis, shape changes and cell division are represented in the CBM and the CPM. For instance, some strengths of the CBM are the ability to simulate large populations and the modeling long-range mechanical effects. However, this comes are the cost of implicit cell shape and less sensitive cell-cell adhesion which renders it less suitable to study the effects of e.g. cell sorting (61). Conversely, the computational costs of the CPM make the simulation of large-scale cellular populations expensive. However, the explicit representation of cell shapes and the cell-cell contacts in CPM make it highly suitable to model tissue dynamics through cell surface mechanics (62).

### 2.6.1 Further reading and software

A detailed review of the full range of cell-based modeling formalisms is presented in (63); see (64) for a more general overview. An interesting comparison between various cell-based modeling formalisms, according to a number of common use cases, is presented in (61).

There is a growing number of dedicated software tools available for cell-based modeling, including several that provide implementations for the cell-based models described in this section. Some platforms such as *PhysiCell* (65) focus on scalability to simulate large-scale populations. Other software tools, most notably *Chaste* (66), are designed for flexibility by providing implementations of various modeling framework within a unified framework, including the ones described in this section. Whereas these software tools typically require substantial programming expertise, the modeling environment *Morpheus* (67) is designed for usability and allows also non-expert to construct multicellular models using a graphical user interface.

## 3 Conclusion

In this chapter, we have introduced a number of statistical and mathematical modeling methods and software tools for the phenotypical analysis of cellular populations. Specifically, we presented methods for the qualitification of cellular behavior in terms of the spatial distribution of cells, their motility, shape and genealogies. Moreover, we have described computational cell-based models that enables one to explore the consequences of hypotheses on the interactions between cells. It should be noted that each of these topics are active fields of research by themselves and this chapter has only scratched the surface of the most commonly used methods and tools. Nevertheless, it is our hope that this chapter will act as a practical guide to statistical and mathematical modeling and will stimulate the reader to engage in suitable quantification, enhancing scientific rigour and possibly leading to significant new biological discoveries.

## Acknowledgements

The work presented in this paper is supported by Deutsche Krebshilfe (SyTASC grant number 70111969) and the German Ministry of Education and Research (BMBF) (HaematoOPT grant number 031A424).

## 4 Notes

1. The locations of cells are, strictly speaking, not independent because cells have a finite volume and cannot overlap (volume exclusion). Therefore, cells will always be more dispersed at short range than expected by CSR.
2. Due to volume exclusion, cellular point patterns will always show dispersion for small radii.
3. In more complex situations, it is possible that the empirical function crosses the theoretical one, e.g. when there is short-range dispersion and long-range clustering.
4. Local (i.e. pointwise) envelopes are calculated for every radius separately, by considering the maximum and minimum value of the simulated K-function. The global envelop is calculated by taking the maximum deviation of all local envelopes, i.e. $L(r) = K_{CSR} - max_r|K(r) - K_{CSR}|$ and $R(r) = K_{CSR} + max_r|K(r) - K_{CSR}|$ where $K_{CSR} = \pi r^2$ in 2D and $K_{CSR} = \frac{4}{3}\pi r^3$ in 3D situations.

5. The time intervals, the time between individual frames, should be identical over the whole trajectory and between trajectories that are being compared.

6. The length of the trajectories, in terms of the number of time intervals, should be as long as possible. Short trajectories may be ignored altogether.

7. Time can be defined in two different ways: When defining time relatively to each trajectory's start, one can detect potential dependencies of the velocity on the cell cycle progress. Conversely, when taking absolute time, one can detect potential temporal dependencies that act globally on all cells in the culture.

8. In case an initial transient is observed followed by constant behavior, one can consider discarding the initial transient time points and limit the analysis for the experiment, excluding the initial transient.

9. Functions for linear regression include `lm` in *R* and `scipy.stats.linregress` in *python*.

10. Typically, cellular trajectories are relatively short. To augment data, one can use overlapping time intervals. E.g. $\tau = 3$ one takes the average over overlapping times $\{t_0, t_2\}$, $\{t_1, t_3\}$, $\{t_2, t_4\}$, etc. instead of the average over $\{t_0, t_2\}$, $\{t_3, t_5\}$, $\{t_6, t_8\}$, etc.

11. Functions for nonlinear regression include `nls` in *R*, and `scipy.optimize.curve_fit` in *python*.

12. For large time intervals, only limited number of samples are averaged over and are therefore unreliable.

13. Analytically derived confidence intervals based on student's t-value are based on the assumption that the data is normal-distributed, which might not be guaranteed. Determining confidence regions by repeated simulation avoids this issue.

14. Most implementations provide axis-aligned bounding boxes, where the bounding box is oriented according to the x- and y-axes of the image instead of the object itself. This implies the statistics are not rotation invariant (diagonal objects have larger bounding box) and results depend on the arbitrary orientation in which they have been imaged. Thus, these properties should be handled with care.

15. PCA is sensitive to the scales of the feature. Therefore, a preprocessing step is necessary to standardize the features. This can be done by subtracting the mean (zero-centering) and dividing by the standard deviation for each column.

16. Inferring correlation structures from cellular genealogies requires highly accurate cell tracking over several generations that so far cannot be achieved by automated tracking methods and thus needs manually created or corrected tracking data.

17. Due to recalculating topological distances, this randomization procedure is computationally expensive. It has to be used e.g. when studying cell death events, since a pure reassigning of death status would lead to genealogies with dead cells having progeny.

# 5 References

1. Krause DS, Scadden DT (2015) A hostel for the hostile: the bone marrow niche in hematologic neoplasms. Haematologica 100 (11):1376-1387.

2. Krinner A, Roeder I (2014) Quantification and Modeling of Stem Cell–Niche Interaction. In: A Systems Biology Approach to Blood. Springer, pp 11-36.

3. Nombela-Arrieta C, Manz MG (2017) Quantification and three-dimensional microanatomical organization of the bone marrow. Blood Adv 1 (6):407-416.

4. Acar M, Kocherlakota KS, Murphy MM, Peyer JG, Oguro H, Inra CN, Jaiyeola C, Zhao Z, Luby-Phelps K, Morrison SJ (2015) Deep imaging of bone marrow shows non-dividing stem cells are mainly perisinusoidal. Nature 526 (7571):126-130.

5. Etzrodt M, Endele M, Schroeder T (2014) Quantitative single-cell approaches to stem cell research. Cell stem cell 15 (5):546-558.

6. Schroeder T (2011) Long-term single-cell imaging of mammalian stem cells. Nature methods 8 (4s):S30.

7. Skylaki S, Hilsenbeck O, Schroeder T (2016) Challenges in long-term imaging and quantification of single-cell dynamics. Nat Biotechnol 34 (11):1137.

8. Foster K, Lassailly F, Anjos-Afonso F, Currie E, Rouault-Pierre K, Bonnet D (2015) Different motile behaviors of human hematopoietic stem versus progenitor cells at the osteoblastic niche. Stem cell reports 5 (5):690-701.

9. Kim S, Lin L, Brown GA, Hosaka K, Scott EW (2017) Extended time-lapse in vivo imaging of tibia bone marrow to visualize dynamic hematopoietic stem cell engraftment. Leukemia 31 (7):1582.

10. Lo Celso C, Lin CP, Scadden DT (2011) In vivo imaging of transplanted hematopoietic stem and progenitor cells in mouse calvarium bone marrow. Nat Protoc 6 (1):1-14.

11. MacLean AL, Smith MA, Liepe J, Sim A, Khorshed R, Rashidi NM, Scherf N, Krinner A, Roeder I, Lo Celso C (2017) Single Cell Phenotyping Reveals Heterogeneity Among Hematopoietic Stem Cells Following Infection. Stem Cells 35 (11):2292-2304.

12. Hilsenbeck O, Schwarzfischer M, Skylaki S, Schauberger B, Hoppe PS, Loeffler D, Kokkaliaris KD, Hastreiter S, Skylaki E, Filipczyk A, Strasser M, Buggenthin F, Feigelman JS, Krumsiek J, van den Berg AJ, Endele M, Etzrodt M, Marr C, Theis FJ, Schroeder T (2016) Software tools for single-cell tracking and quantification of cellular and molecular properties. Nat Biotechnol 34 (7):703-706.

13. Hilsenbeck O, Schwarzfischer M, Loeffler D, Dimopoulos S, Hastreiter S, Marr C, Theis FJ, Schroeder T (2017) fastER: a user-friendly tool for ultrafast and robust cell segmentation in large-scale microscopy. Bioinformatics 33 (13):2020-2028.

14. Molnar C, Jermyn IH, Kato Z, Rahkama V, Östling P, Mikkonen P, Pietiäinen V, Horvath P (2016) Accurate morphology preserving segmentation of overlapping cells based on active contours. Sci Rep 6:32412.

15. Sommer C, Straehle C, Koethe U, Hamprecht FA Ilastik: Interactive learning and segmentation toolkit. In: Biomedical Imaging: From Nano to Macro, 2011 IEEE International Symposium on, 2011. IEEE, pp 230-233.

16. Pelt DM, Sethian JA (2018) A mixed-scale dense convolutional neural network for image analysis. Proc Natl Acad Sci U S A 115 (2):254-259.

17. Ronneberger O, Fischer P, Brox T U-net: Convolutional networks for biomedical image segmentation. In: International Conference on Medical image computing and computer-assisted intervention, 2015. Springer, pp 234-241.

18. Meijering E (2012) Cell segmentation: 50 years down the road [life sciences]. IEEE Signal Processing Magazine 29 (5):140-145.

19. Kan A (2017) Machine learning applications in cell image analysis. Immunol Cell Biol 95 (6):525-530.

20. Caicedo JC, Cooper S, Heigwer F, Warchal S, Qiu P, Molnar C, Vasilevich AS, Barry JD, Bansal HS, Kraus O (2017) Data-analysis strategies for image-based cell profiling. Nature methods 14 (9):849.

21. Schindelin J, Arganda-Carreras I, Frise E, Kaynig V, Longair M, Pietzsch T, Preibisch S, Rueden C, Saalfeld S, Schmid B (2012) Fiji: an open-source platform for biological-image analysis. Nature methods 9 (7):676.

22. Carpenter AE, Jones TR, Lamprecht MR, Clarke C, Kang IH, Friman O, Guertin DA, Chang JH, Lindquist RA, Moffat J (2006) CellProfiler: image analysis software for identifying and quantifying cell phenotypes. Genome Biol 7 (10):R100.

23. Held M, Schmitz MH, Fischer B, Walter T, Neumann B, Olma MH, Peter M, Ellenberg J, Gerlich DW (2010) CellCognition: time-resolved phenotype annotation in high-throughput live cell imaging. Nature methods 7 (9):747.

24. Tinevez J-Y, Perry N, Schindelin J, Hoopes GM, Reynolds GD, Laplantine E, Bednarek SY, Shorte SL, Eliceiri KW (2017) TrackMate: An open and extensible platform for single-particle tracking. Methods 115:80-90.

25. Wiesmann V, Franz D, Held C, Münzenmayer C, Palmisano R, Wittenberg T (2015) Review of free software tools for image analysis of fluorescence cell micrographs. J Microsc 257 (1):39-53.

26. Nilsson SK, Johnston HM, Coverdale JA (2001) Spatial localization of transplanted hemopoietic stem cells: inferences for the localization of stem cell niches. Blood 97 (8):2293-2299.

27. Ripley BD (1976) The second-order analysis of stationary point processes. J Appl Probab 13 (2):255-266.

28. Baddeley A (1999) Spatial sampling and censoring. In: Barndorff-Nielsen O, Kendall W, van Lieshout H (eds) Stochastic geometry: likelihood and computation. Chapman and Hall (London), pp 37-78.

29. Baddeley A, Rubak E, Turner R (2015) Spatial point patterns: methodology and applications with R. CRC Press,

30. Cressie N (2015) Statistics for spatial data. John Wiley & Sons,

31. Gelfand AE, Diggle P, Guttorp P, Fuentes M (2010) Handbook of spatial statistics. CRC press,

32. Tranquillo RT, Lauffenburger DA, Zigmond S (1988) A stochastic model for leukocyte random motility and chemotaxis based on receptor binding fluctuations. The Journal of cell biology 106 (2):303-309.

33. Wu P-H, Giri A, Sun SX, Wirtz D (2014) Three-dimensional cell migration does not follow a random walk. Proceedings of the National Academy of Sciences 111 (11):3949-3954.

34. Luzhanskey ID, MacMunn JP, Cohen JD, Barney LE, Jansen LE, Schwartz AD, Peyton S (2017) Anomalous Diffusion as a Descriptive Model of Cell Migration. bioRxiv:236356.

35. Gorelik R, Gautreau A (2014) Quantitative and unbiased analysis of directional persistence in cell migration. Nat Protoc 9 (8):1931.

36. Wu P-H, Giri A, Wirtz D (2015) Statistical analysis of cell migration in 3D using the anisotropic persistent random walk model. Nat Protoc 10 (3):517.

37. Dieterich P, Klages R, Preuss R, Schwab A (2008) Anomalous dynamics of cell migration. Proceedings of the National Academy of Sciences 105 (2):459-463.

38. Makarava N, Menz S, Theves M, Huisinga W, Beta C, Holschneider M (2014) Quantifying the degree of persistence in random amoeboid motion based on the Hurst exponent of fractional Brownian motion. Physical Review E 90 (4):042703.

39. Masuzzo P, Van Troys M, Ampe C, Martens L (2016) Taking aim at moving targets in computational cell migration. Trends Cell Biol 26 (2):88-110.

40. Sánchez-Corrales YE, Hartley M, van Rooij J, Marée AF, Grieneisen VA (2018) Morphometrics of complex cell shapes: lobe contribution elliptic Fourier analysis (LOCO-EFA). Development:dev. 156778.

41. Pincus Z, Theriot J (2007) Comparison of quantitative methods for cell-shape analysis. J Microsc 227 (2):140-156.

42. Driscoll MK, McCann C, Kopace R, Homan T, Fourkas JT, Parent C, Losert W (2012) Cell shape dynamics: from waves to migration. PLoS Comput Biol 8 (3):e1002392.

43. Gordonov S, Hwang MK, Wells A, Gertler FB, Lauffenburger DA, Bathe M (2016) Time series modeling of live-cell shape dynamics for image-based phenotypic profiling. Integrative Biology 8 (1):73-90.

44. Glauche I, Lorenz R, Hasenclever D, Roeder I (2009) A novel view on stem cell development: analysing the shape of cellular genealogies. Cell Prolif 42 (2):248-263.

45. Bach E, Zerjatke T, Herklotz M, Scherf N, Niederwieser D, Roeder I, Pompe T, Cross M, Glauche I (2014) Elucidating functional heterogeneity in hematopoietic progenitor cells: a combined experimental and modeling approach. Exp Hematol 42 (9):826-837 e821-817.

46. Khakhutskyy V, Schwarzfischer M, Hubig N, Plant C, Marr C, Rieger MA, Schroeder T, Theis FJ Centroid Clustering of Cellular Lineage Trees. In: International Conference on Information Technology in Bio-and Medical Informatics, 2014. Springer, pp 15-29.

47. Stadler T, Skylaki S, K DK, Schroeder T (2018) On the statistical analysis of single cell lineage trees. J Theor Biol 439:160-165.

48. Marr C, Strasser M, Schwarzfischer M, Schroeder T, Theis FJ (2012) Multi-scale modeling of GMP differentiation based on single-cell genealogies. FEBS J 279 (18):3488-3500.

49. Nordon RE, Ko K-H, Odell R, Schroeder T (2011) Multi-type branching models to describe cell differentiation programs. J Theor Biol 277 (1):7-18.

50. Strasser MK, Feigelman J, Theis FJ, Marr C (2015) Inference of spatiotemporal effects on cellular state transitions from time-lapse microscopy. BMC Syst Biol 9 (1):61.

51. Feigelman J, Ganscha S, Hastreiter S, Schwarzfischer M, Filipczyk A, Schroeder T, Theis FJ, Marr C, Claassen M (2016) Analysis of cell lineage trees by exact Bayesian inference identifies negative autoregulation of Nanog in mouse embryonic stem cells. Cell systems 3 (5):480-490. e413.

52. d'Inverno M, Luck M, Luck MM (2004) Understanding agent systems. Springer Science & Business Media,

53. Krinner A, Roeder I, Loeffler M, Scholz M (2013) Merging concepts-coupling an agent-based model of hematopoietic stem cells with an ODE model of granulopoiesis. BMC Syst Biol 7 (1):117.

54. Roeder I, Horn M, Glauche I, Hochhaus A, Mueller MC, Loeffler M (2006) Dynamic modeling of imatinib-treated chronic myeloid leukemia: functional insights and clinical implications. Nat Med 12 (10):1181.

55. Deutsch A, Dormann S (2007) Cellular automaton modeling of biological pattern formation: characterization, applications, and analysis. Springer Science & Business Media,

56. Graner F, Glazier JA (1992) Simulation of biological cell sorting using a two-dimensional extended Potts model. Phys Rev Lett 69 (13):2013.

57. Drasdo D (2007) Center-based single-cell models: An approach to multi-cellular organization based on a conceptual analogy to colloidal particles. In: Single-Cell-Based Models in Biology and Medicine. Springer, pp 171-196.

58. Alt S, Ganguly P, Salbreux G (2017) Vertex models: from cell mechanics to tissue morphogenesis. Phil Trans R Soc B 372 (1720):20150520.

59. Fletcher AG, Osterfield M, Baker RE, Shvartsman SY (2014) Vertex models of epithelial morphogenesis. Biophys J 106 (11):2291-2304.

60. Sandersius SA, Newman TJ (2008) Modeling cell rheology with the subcellular element model. Phys Biol 5 (1):015002.

61. Osborne JM, Fletcher AG, Pitt-Francis JM, Maini PK, Gavaghan DJ (2017) Comparing individual-based approaches to modelling the self-organization of multicellular tissues. PLoS Comput Biol 13 (2):e1005387.

62. Magno R, Grieneisen VA, Marée AF (2015) The biophysical nature of cells: potential cell behaviours revealed by analytical and computational studies of cell surface mechanics. BMC Biophys 8 (1):8.

63. Van Liedekerke P, Palm M, Jagiella N, Drasdo D (2015) Simulating tissue mechanics with agent-based models: concepts, perspectives and some novel results. Computational Particle Mechanics 2 (4):401-444.

64. Tanaka S (2015) Simulation frameworks for morphogenetic problems. Computation 3 (2):197-221.

65. Ghaffarizadeh A, Heiland R, Friedman SH, Mumenthaler SM, Macklin P (2018) PhysiCell: An open source physics-based cell simulator for 3-D multicellular systems. PLoS Comput Biol 14 (2):e1005991.

66. Mirams GR, Arthurs CJ, Bernabeu MO, Bordas R, Cooper J, Corrias A, Davit Y, Dunn S-J, Fletcher AG, Harvey DG (2013) Chaste: an open source C++ library for computational physiology and biology. PLoS Comput Biol 9 (3):e2970.

67. Starruß J, de Back W, Brusch L, Deutsch A (2014) Morpheus: a user-friendly modeling environment for multiscale and multicellular systems biology. Bioinformatics 30 (9):1331-1332.